\documentclass{ifacconf}

\usepackage{graphicx}      
\usepackage{natbib}        
\usepackage{booktabs}
\usepackage{bm}
\usepackage{fp, tikz, tikz-cd, pgfplots}

\usepackage{algorithmic}
\usepackage{mathtools}

\usepackage{array}
\usepackage{amsmath, amssymb, natbib, graphicx, url}
\usepackage{setspace}

\usepackage{tablefootnote}

\usetikzlibrary{arrows}
\usetikzlibrary{shapes}
\usetikzlibrary{patterns}
\usetikzlibrary{decorations.pathmorphing}
\usetikzlibrary{decorations.pathreplacing}
\usetikzlibrary{decorations.shapes}
\usetikzlibrary{decorations.text}
\usetikzlibrary{shapes.misc}
\usetikzlibrary{decorations.markings}
\usetikzlibrary{arrows.meta}
\usetikzlibrary{backgrounds}
\usepgfplotslibrary{fillbetween}
\usepgfplotslibrary{statistics}

\usetikzlibrary{shapes,arrows,fit,backgrounds,positioning}
\usetikzlibrary{decorations.pathmorphing} 
\usetikzlibrary{matrix} 
\usetikzlibrary{arrows} 
\usetikzlibrary{calc} 
\usepackage{verbatim}

\usepackage{pgfplots}
\DeclareUnicodeCharacter{2212}{−}
\usepgfplotslibrary{groupplots,dateplot}
\usetikzlibrary{patterns,shapes.arrows}
\pgfplotsset{compat=newest}

\newif\ifarxiv
\arxivtrue

\newtheorem{assumption}{Assumption}

\newtheorem{proposition}{Proposition}

\newtheorem{definition}{Definition}
\newtheorem{theorem}{Theorem}
\newtheorem{remark}{Remark}
\newtheorem{example}{Example}

\makeatletter
\def\blfootnote{\gdef\@thefnmark{}\@footnotetext}
\makeatother

\newcommand\clr[2]{{\color{#1}{#2}}}
\newcommand\ko[0]{{\mathcal{K}}}

\newcommand\Set[1]{\mathbb{#1}} 
\newcommand{\tsgn}[1]{{#1}}

\begin{document}
    \newlength\fheight 
    \newlength\fwidth 
\begin{frontmatter}

\title{Towards Data-driven LQR with {Koopmanizing Flows}$^{\star}$} 

\thanks[footnoteinfo]{\scriptsize This work was supported by the European Union's Horizon 2020 research and innovation programme under grant agreement no. 871295 "SeaClear".}

\author[First]{Petar Bevanda} 
\author[First]{Max Beier} 
\author[Second]{Shahab Heshmati-Alamdari}
\author[First]{Stefan Sosnowski}
\author[First]{Sandra Hirche}
\address[First]{Chair of Information-oriented Control (ITR), Department of Electrical and Computer Engineering,
Technical University of Munich
D-80333 Munich, Germany (e-mail: $[$petar.bevanda, max.beier, sosnowski, hirche$]$@tum.de).}
\address[Second]{Department of Electronic Systems, 
  Aalborg University, Fredrik Bajers Vej 7K, 9220 Aalborg,
Denmark (e-mail: shhe@es.aau.dk)}

\begin{abstract}                
{We propose a novel framework for learning linear time-invariant (LTI) models for a class of continuous-time non-autonomous nonlinear dynamics based on a representation of Koopman operators. In
general, the operator is infinite-dimensional but, crucially, linear. To utilize it for efficient LTI control design, we learn a finite representation of the Koopman operator that is linear in controls while concurrently learning meaningful lifting coordinates. For the latter, we rely on \textit{Koopmanizing Flows} - a diffeomorphism-based representation of Koopman operators {and extend it to systems with linear control entry.}
With such a learned model, we can replace the nonlinear optimal control problem with quadratic cost to that of a linear quadratic regulator (LQR), facilitating efficacious optimal control for nonlinear systems.  The superior control performance of the proposed method is demonstrated on simulation examples.}
\end{abstract}


\end{frontmatter}
\looseness=-1
\section{Introduction}
Inspired by the infinite-dimensional, \textit{linear} Koopman operator \citep{Koopman1931}, there has been an increased interest in global linearizations of nonlinear dynamics in recent years. By taking a finite-dimensional nonlinear system and ``lifting" it to a higher-dimensional linear operator representation, superior complexity-accuracy balance compared to conventional nonlinear modeling is possible via efficient linear techniques for prediction and control.\blfootnote{\scriptsize \textcopyright 2022 the authors. Accepted by IFAC for publication under a Creative Commons Licence CC-BY-NC-ND.}

Although the autonomous setting is native to the Koopman operator, there are practicable extensions of the theory to controlled systems that open up the possibility of applying classical results from linear control theory to nonlinear systems \citep{Bevanda2021a}.
First ideas of including control in Koopman-inspired frameworks can be found in \cite{doi:10.1137/16M1062296}, with ideas of optimal control analytically considered in \cite{Brunton2016a} and a data-driven manner for reduced order optimal control in \cite{Kaiser2021}.
Yet, the aforementioned works are not geared towards linear predictors - leading to an unclear relation between the Koopman-inspired reformulation of the nonlinear optimal control problem.  Other existing optimal control methods based on data-driven linear predictors utilize linear model predictive control (MPC) \citep{Korda2018a,Lian2020}, inspiring control applications in nonlinear flows \citep{Arbabi2017}, soft robotics \citep{Bruder2020} and autonomous vehicles \citep{Cibulka2020}.
Often, however, the aforementioned data-driven approaches tend to provide only locally accurate models and operate in a receding-horizon fashion.
However, a data-driven solution of an optimal control problem for nonlinear systems using linear techniques might be more attractive due to the possibility of solving it in closed form.
To exploit Koopman operator representations advantageously through an LQR problem reformulation, it is crucial to obtain linear predictors that are long-term accurate - something not previously explored.
{Previous works demonstrate long-term accurate autonomous prediction when learning LTI features for nonlinear systems with prior spectral knowledge of the operator \citep{Bevanda2021b} or learn LTI features and spectra concurrently \citep{Bevanda2021c}. Crucially, none consider actuated systems and closed-form optimal control.}
Inspired by the prediction efficacy and theoretical properties of \textit{Koopmanizing Flows} \citep{Bevanda2021c}, we propose an extension to controlled systems allowing for efficient LQR control design. The contribution of this paper is the development of {\textit{Koopmanizing Flows with Control LQR} (KFc-LQR)} - a {principled} framework for learning controlled Koopman operator dynamical models admitting an LQR design for nonlinear optimal control.
The framework is entirely data-driven as the lifting and the controlled LTI dynamics are learned concurrently.
Such a learned model allows us to cast a nonlinear  optimal control problem with quadratic cost to that of an LQR.
To the best of our knowledge, this is the first Koopman-based framework for fully data-driven LQR control design. We demonstrate the superior performance of KFc-LQR on simulation examples.

\section*{Notation}
		Vectors/matrices are denoted with lower/upper case bold symbols $\bm{x}$/$\bm{X}$. Symbols $\mathbb{N}/\mathbb{ R }/\mathbb{C}$ denote sets of natural/real/complex numbers, while $\mathbb{N}_{0}$ denotes all natural numbers with zero, and $\mathbb{R}_{+,0}/\mathbb{R}_{+}$ all positive reals with/without zero. 
		Function spaces with a specific integrability/regularity order are denoted as $L^{}$/$C^{}$ with the order (class) specified in their exponent. The Jacobian matrix of vector-valued map $\bm{\psi}$ evaluated at $\bm{x}$ is denoted as $\bm{J}_{\bm{\psi}}(\bm{x})$.
The $L^p$-norm on a set $\Set{X}$ is denoted as $\|\!\cdot\!\|_{p, \Set{X}}$. Writing $\odot$ denotes the Hadamard product, $\operatorname{exp}$ pointwise exponential and $\circ$ function composition. Underlined matrices $\underline{\bm{X}}$ represent ones in the immediate (original) state-space.
\section{Problem Statement}
\subsection{Modeling Assumptions}\label{mod:ass}
Consider an unknown, continuous-time nonlinear dynamical system of the following form 
	\begin{equation}\label{eq:sys}
	\dot{\bm{x}}=\bm{a}(\bm{x})+\underline{\bm{B}}\bm{u}=:\bm{f}(\bm{x},\bm{u})
	\end{equation}
 with continuous states and controls on compact sets $\bm{x} \in \Set{X} \subset \mathbb{R}^{d}$ and $\boldsymbol{u} \in \mathbb{U} \subset \mathbb{R}^{m}$, respectively. The autonomous dynamics are smooth such that $\bm{a} \in C^2(\Set{X})$, with control assumed to be entering the dynamics linearly via $\underline{\bm{B}} \in \mathbb{R}^{ d \times m}$.
\begin{assumption}\label{ass:sysCLS}
We assume the sole fixed point of $\bm{f}(\bm{x},\bm{0})$ for \eqref{eq:sys} is hyperbolic and contained in $\mathbb{X}$.
\end{assumption}
The above assumption admits dynamical systems representing motion, e.g., human reaching movements \citep{Khansari2011} or physical systems such as a neutrally buoyant underwater vehicle \citep{bookFOSS}.

\clr{blue}{}
\subsection{Koopman Operator Theory}
The solutions of forward-complete continuous-time dynamics \citep{Bittracher2015} are fully described by the flow map of $\bm{a}(\bm{x}):=\bm{f}(\bm{x},\bm{0})$ 
\begin{equation}
\label{flow}
 \boldsymbol{x}(t_0) \equiv \boldsymbol{x}_0, \quad \boldsymbol{F}^{t}(\boldsymbol{x}_0):= \boldsymbol{x}_0+\int_{t_{0}}^{t_{0}+t} \boldsymbol{a}(\boldsymbol{x}(\tau)) d \tau,
\end{equation}
which has a unique solution on $[0,+\infty)$ from the initial condition $\boldsymbol{x}$ at $t = 0$ ~\citep{Angeli1999}. Due to the hyperbolicity of the isolated attractor, this holds for \eqref{eq:sys}. The above flow map induces the associated Koopman operator semigroup as defined in the following.
\begin{definition}\label{def:Koop}
	The semigroup $\{{\mathcal{K}}^{t}\}_{t \in \mathbb{R}_{+,0}}\!:\! C(\Set{X}) \!\mapsto\! C(\Set{X})$ of  Koopman operators for the autonomous flow of \eqref{flow} acts on a scalar observable function ${h} \!\in\! C(\Set{X})$ on the state space $\Set{X}$ through ${\mathcal{K}^{t}_{\boldsymbol{a}}} {{h}} ={{h}}\circ{\boldsymbol{F}^{t}}$.
\end{definition}
Simply put, the operator applied to an observable function $h$ at time $t_0$ advances it along the flow \eqref{flow} as follows $\ko^t_{\bm{a}}{h}(\bm{x}(t_0))={h}(\bm{x}(t_0+t))$. By applying it component-wise to the ``state-observer" function $\bm{h}(\bm{x})=\bm{x}$, it is identical to the flow map defined in \eqref{flow}. Critically, every $\ko^{t}_{\bm{a}}$ is a \textit{linear}\footnote{Consider $h_{1}, h_{2} \in {C}(\Set{X})$ and $\beta \in \Set{C}$. Then, using Definition \ref{def:Koop}, $\mathcal{K}^{t}\left(\beta h_{1}+h_{2}\right)=\left(\beta h_{1}+h_{2}\right) \circ \bm{F}^{t}=\beta h_{1} \circ \bm{F}^{t}+h_{2} \circ \bm{F}^{t}=$ $\beta \mathcal{K}^{t} h_{1}+\mathcal{K}^{t} h_{2}$.} operator.
With a well-defined Koopman operator semigroup, we introduce its infinitesimal generator.
\begin{definition}\label{def:generator}
	The evolution operator
	\begin{equation}
	\mathcal{G}_{{\mathcal{K}_{\bm{a}}}} {h}=\lim _{t \rightarrow 0^{+}} \frac{{\mathcal{K}}^{t}_{\bm{a}} {h}-{h}}{t} = \frac{d}{dt}{h},
	\end{equation}
	is the infinitesimal generator of the semigroup of Koopman operators $\{{\mathcal{K}}^{t}\}_{t \in \mathbb{R}_{+,0}}$.
\end{definition}
Crucially, the Koopman operator formalism allows one to decompose dynamics into linearly evolving coordinates, which naturally arise through the eigenfunctions of evolution operator $\mathcal{G}_{{\mathcal{K}_{\bm{a}}}}$. These eigenfunctions are formally defined as follows.
\begin{definition}\label{eigF}
	An observable $\phi \in C(\Set{X})$ is an \textit{eigenfunction} of $\mathcal{G}_{{\mathcal{K}_{\bm{a}}}}$ if it satisfies $\mathcal{G}_{\mathcal{K}_{\bm{a}}} \phi = \lambda \phi$,
	for an \textit{eigenvalue} $\lambda \in \mathbb{C}$. The span of eigenfunctions $\phi$ of $\mathcal{G}_{{\mathcal{K}_{\bm{a}}}}$ is denoted by $\bm{\Phi}$.
\end{definition}
Due to Assumption~\ref{ass:sysCLS}, the Koopman operator generator has a pure point spectrum for the dynamics \eqref{eq:sys} \citep{Mauroy2016b}. 
Therefore, for each observable $h$, there exists a sequence $v_j(h)\in\Set{C}$ of mode weights, such that the following decomposition \emph{completely} describes its dynamics as 
\begin{equation}\label{eq:KMD}
\dot{h} =\mathcal{G}_{\ko_{\bm{a}}} h =\mathcal{G}_{\ko_{\bm{a}}}\left(\sum_{j=1}^{\infty} v_{j}(h) {\phi}_{j}\right)=\sum_{j=1}^{\infty} v_{j}(h) \lambda_{j} {\phi}_{j}.
\end{equation}
With a slight abuse of operator notation, we can write the decomposition \eqref{eq:KMD} compactly as $\dot{h}=\mathcal{V}_{h}\mathcal{G}_{\ko_{\bm{a}}}\bm{\Phi}$, where $\mathcal{V}_{h}$ is an operator projecting on the observable.

The above decomposition describes autonomous dynamics, native to Koopman operators - requiring extensions for control systems. While there are various options for handling control dynamics, we side with one that is practical and covers the class of systems considered in \eqref{eq:sys}.
We extend \eqref{eq:KMD} by additive input-linear terms with $\bm{b}_j \in \mathbb{R}^{m}$ such that
\begin{equation}\label{eq:KMDc}
\dot{h} =\mathcal{G}_{\ko_{\bm{a}}}\left(\sum_{j=1}^{\infty} v_{j}(h) {\phi}_{j}\right)=\sum_{j=1}^{\infty} v_{j}(h) \lambda_{j} {\phi}_{j}+\sum_{j=1}^{\infty} v_{j}(h) \bm{b}_{j}^{\top} \bm{u},
\end{equation}
holds for all systems of type \eqref{eq:sys}.
This is easily checked by considering an additional linear operator $\mathcal{B}$ in the following operator-based description of \eqref{eq:KMDc}:
\begin{equation}\label{eq:KMDcOp}
    \dot{h}=\mathcal{V}_{h}\mathcal{G}_{\ko_{\bm{a}}}\bm{\Phi}+\mathcal{V}_{h}\mathcal{B}\bm{u}
\end{equation}
The autonomous part of \eqref{eq:KMDcOp} satisfies \eqref{eq:KMD} and the additive control part remains linear as in \eqref{eq:sys}.
\begin{remark}
If the actuation entry in \eqref{eq:sys} was nonlinear, assuming an LTI controlled system in the lifted model becomes only locally accurate \citep{Bevanda2021a} but still can suffice for short-horizon prediction and control \citep{Korda2018a,Lian2020}.
\end{remark}

Clearly, the infinite-dimensional model as in \eqref{eq:KMDcOp} is not helpful for a practical representation. As finding meaningful finite-dimensional models for \eqref{eq:KMDcOp} is generally might not analytically feasible even with perfect system knowledge, we seek a finite-dimensional representation from data.  Note that, due to the linearity of input entry, it need not be lifted but projected onto the LTI system of the autonomous dynamics.

\begin{assumption}\label{ass:data}
A data-set of $N$ state-input-output tuples $\mathbb{D}_{N}=\left\{\bm{x}^{(i)}, \bm{u}^{(i)},\dot{\bm{x}}^{(i)}\right\}_{i=1}^{N}$ for the system~\eqref{eq:sys} 
is available.
\end{assumption} 
The above measurements are commonly assumed to be at disposal. If not directly accessible, the time-derivative of the state can be approximated through finite differences for practical applications.
Based on the data from Assumption \ref{ass:data}, we consider the problem of learning a finite-dimensional Koopman generator model for the forced system \eqref{eq:sys} by solving the following optimization problem
\begin{subequations}\label{opt_prob}
    \begin{gather}
      \begin{aligned}
\operatornamewithlimits{min}_{\bm{{A}}, \bm{B},\bm{C}, \boldsymbol{\psi}(\cdot)}  \sum^{N}_{i=1}&\overbrace{\|\dot{\bm{x}}^{(i)}-\bm{C}\left(\bm{A}\boldsymbol{\psi}(\bm{x}^{(i)})+\bm{B}\bm{u}^{(i)}\right)\|^{2}_2}^{\text{prediction}} \\&+\overbrace{\|\bm{x}^{(i)}-\boldsymbol{C}\boldsymbol{\psi}(\bm{x}^{(i)})\|^{2}_2}^{\text{reconstruction}} 
      \end{aligned} \label{obj_opt} 
    \end{gather}
\begin{align}
\text{subject to:} &  \quad 
\boldsymbol{\psi} \in \bm{\Phi}  \quad \quad \text{(Koopman-invariance)} \label{space_opt}
\end{align}
\end{subequations}
with $\bm{\psi}=[\psi_1,\cdots,\psi_D]^{\top}$, $\bm{{A}} \in \Set{R}^{D \times D}$, $\bm{{B}} \in \Set{R}^{D \times m}$ and $\bm{C} \in \Set{R}^{d \times D}$ providing a finite-dimensional representation in terms of a state-space model
\begin{subequations}\label{eq:LTI full}
\begin{align}
\bm{z}^{}_0 &=\bm{\psi}^{}(\bm{x}_0), \label{eq:LTI:1}\\
\dot{\bm{z}}^{} &=\bm{{A}}\bm{z}^{}+\bm{{B}}\bm{u}^{}, \label{eq:LTI:2}\\
\hat{\bm{x}} &=\bm{C} \bm{z}^{}. \label{eq:LTI:3}
\end{align}
\end{subequations}
This model trades the nonlinearity of a $d$-dimensional ODE (\ref{eq:sys}) for a nonlinear ``lift" (\ref{eq:LTI:1}) of the initial condition $\bm{x}_0$ to higher dimensional ($D \gg d$) Koopman-invariant coordinates \eqref{eq:LTI:2} such that the original state can be linearly reconstructed via \eqref{eq:LTI:3}. Moreover, \eqref{opt_prob} is geared towards learning an arbitrary amount of Koopman-invariant features directly instead of only finding ones that lie in a heuristically predetermined dictionary of functions.

\section{Learning a Controlled Diffeomorphism-based LTI Model}
To preface the learning approach, we formally define the notion of LTI-coordinates - ones that evolve linearly under the dynamics.
\begin{definition}\label{lem:LTIbase}
Consider the system \eqref{eq:sys}, a matrix $\bm{{A}} \in \Set{R}^{D \times D}$ and a finite set of features $\bm{\psi}:=\left[\psi_{1}(\bm{x}), \ldots, \psi_{D}(\bm{x})\right]^{\top}$ with $\psi_i(\bm{x}) \in C^{1}(\Set{X})$ on a compact set $\mathbb{X}$. If this feature set solves the following linear partial differential equation (PDE)
\begin{equation}\label{eq:PDElift}
\bm{J}_{\bm{\psi}}(\bm{x})\bm{a}(\bm{x})=\bm{\bm{{A}}} \bm{\psi}(\bm{x}),
\end{equation}
the features are admissible Koopman-invariant coordinates satisfying \eqref{space_opt}.
\end{definition} 

\begin{figure}[t!]
    \centering
  \begin{tikzpicture}[baseline= (a).base, ampersand replacement=\&]
    \node[scale=1.0] (a) at (0,0){
    	\begin{tikzcd}[column sep = huge, row sep = huge, ]
    	\mathbb{X} \arrow{r}{\color{magenta}\boldsymbol{d}} \arrow[swap]{d}{\boldsymbol{a}} \arrow{d}{}  \arrow[thick, bend left]{rr}{\bm{\psi}} \arrow[thick,swap, bend left]{rr}{} \& \mathbb{Y}_{\Box} \arrow[swap]{d}{\color{magenta}\underline{\bm{A}}} \arrow{r}{\bm{\varrho}:~\bm{y} \mapsto \bm{y}^{[\overline{p}]}} \arrow[swap]{r}{} \& \mathbb{Z}  
    	\arrow[swap,thick]{d}{\bm{{\boldsymbol{A}}_{[\overline{p}]}}({\color{magenta} \underline{\bm{A}}})}\\%
    	\mathcal{T}_{\bm{x}}\mathbb{X} \arrow{r}{\frac{\partial \color{magenta}\bm{d}}{\partial \bm{x}}} \& \mathcal{T}_{\bm{y}}\mathbb{Y}_{\Box} \& \mathcal{T}_{\bm{z}}\mathbb{Z}
    	\arrow[thick,swap, bend left, dashrightarrow]{ll}{\color{magenta}\bm{C}}
    	\end{tikzcd}
    };
    \end{tikzpicture}    
    \vspace{-0.4cm}
    \caption{Construction diagram for learning model for the autonomous part of \eqref{eq:LTI:1}-\eqref{eq:LTI:3} with the construction pathway in bold and the maps to be learned in magenta. The sets $\Set{X},\Set{Y}_{\Box},\Set{Z}$ correspond to the immediate state-space, latent unit-box space and lifted linear model space, respectively; with corresponding tangent spaces denotes as $\mathcal{T}_{\bm{x}}\Set{X},\mathcal{T}_{\bm{y}}\Set{Y}_{\Box},\mathcal{T}_{\bm{z}}\Set{Z}$.
    }
    \label{fig:commDiag}
\end{figure}
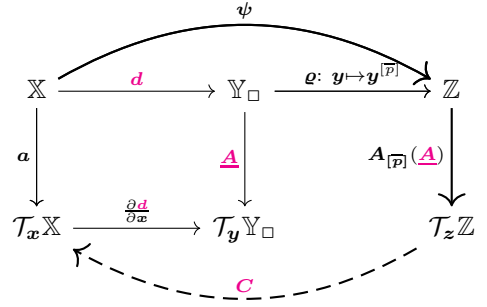

\subsection{Structured Lifting Construction from a Latent Space}

As in \textit{Koopmanizing Flows} \citep{Bevanda2021c}, we use a diffeomorphic relation to a latent linear model to obtain solutions for \eqref{eq:PDElift}, providing us with Koopman-invariant lifting coordinates that fulfill \eqref{space_opt}. The \textit{Koopmanizing Flows} framework relies on the fact that lifting a linear systems to a monomial basis is invariant to the dynamics, meaning the monomial coordinates still form an LTI system whose spectral properties are determined by the original linear system itself. To exploit the former for nonlinear system modeling, one is tasked with morphing the nonlinear system to a linear one by a smooth change of coordinates.
\begin{definition}\label{def:smthEQ}
    Vector fields $\dot{\bm{x}}=\bm{a}(\bm{x})$ and $\dot{\bm{y}}=\bm{t}(\bm{y})$ are diffeomorphic, or smoothly equivalent, if there exists a diffeomorphism $\bm{d}: \mathbb{R}^{d} \mapsto \mathbb{R}^{d}$ such that $ \bm{a}(\bm{x})=\bm{J}^{-1}_{\bm{d}}(\bm{x})\bm{t}(\bm{d}(\bm{x}))$.
\end{definition}

We first formally define monomial coordinates based on the latent vector  $\bm{d}(\bm{x})=\bm{y}=[y_1,\dots,y_d]^{\top}$ through $y^{\bm{\alpha}}=y_{1}^{\alpha_{1}} y_{2}^{\alpha_{2}} \cdots y_{d}^{\alpha_{d}}$, 
where $\bm{\alpha}\in\mathbb{N}_0^d$ is a multi-index. 
Then, we obtain a lifted coordinate vector by concatenating all monomials $y^{\bm{\alpha}}$ up to order $\|\bm{\alpha}\|_1=\alpha_{1}+\cdots+\alpha_{d}\leq\overline{p}$ in a lexicographical ordering in a vector $\bm{y}^{[\overline{p}]}$. Due to the construction of this vector, it inherits the linear dynamical system description from $\bm{y}=\bm{d}(\bm{x})$ with the dynamics of $\bm{y}^{[\overline{p}]}$ linearly dependent on $\underline{\bm{A}}$ \citep[Lemma 6]{Bevanda2021c}. This is formalized in the following proposition.
\begin{proposition}[\cite{Bevanda2021c}]\label{prop1}
Assume the linear system $\dot{\bm{y}}=\underline{\bm{A}}\bm{y}$ is smoothly equivalent to the autonomous part of ~\eqref{eq:sys} via a diffeomorphism $\bm{d}$ such that ${\bm{y}}=\bm{d}(\bm{x})$. Then, the lifted features $\bm{\psi}=\bm{d}^{{[\overline{p}]}}$ satisfy \eqref{space_opt}, i.e.,  $\bm{\psi}(\bm{x})=\bm{d}^{{[\overline{p}]}}(\bm{x})=\bm{y}^{{[\overline{p}]}}$ are Koopman-invariant coordinates and define a latent linear system\looseness=-1
\begin{subequations}\label{eq:liftedLTI full}
\begin{align}
\bm{z}^{}_0 &=\bm{d}^{{[\overline{p}]}}(\bm{x}_0), \label{eq:liftedLTI:1}\\
\dot{\bm{z}}^{} &=\bm{A}_{[\overline{p}]}(\underline{\bm{A}})\bm{z}^{}. \label{eq:liftedLTI:2}
\end{align}
\end{subequations}
\end{proposition}
Due the above result, learning Koopman-invariant features reduces to learning a diffeomorphism, which allows us to replace the constraint \eqref{space_opt} by the condition 
\begin{align}\label{eq:smooth_equi}
    \bm{a}(\bm{x})=\bm{J}^{-1}_{\bm{d}}(\bm{x})\underline{\bm{A}}\bm{d}(\bm{x}),
\end{align}
which ensures the smooth equivalence between $\bm{a}$ and $\underline{\bm{A}}$.

\begin{theorem}\label{thm:singleConstr}
The minimizers $\underline{\hat{\bm{A}}}, \hat{\bm{B}}, \hat{\bm{C}}, \hat{\bm{\bm{d}}}(\cdot)$ of the optimization problem
\begin{subequations}\label{eq:step2opt}
    \begin{gather}
      \begin{aligned}
\operatornamewithlimits{min}_{\substack{\underline{\bm{A}},\bm{B}, \bm{C}, \bm{\bm{d}}(\cdot)}} \sum^{N}_{i=1} &\|\dot{\bm{x}}^{(i)}-\bm{C}(\bm{A}_{[\overline{p}]}(\underline{\bm{A}})\bm{d}^{{[\overline{p}]}}(\bm{x}^{(i)})+\bm{B}\bm{u}^{(i)})\|^{2}_2 \\ + &\|\bm{x}^{(i)}-\boldsymbol{C}\bm{d}^{{[\overline{p}]}}(\bm{x}^{(i)})\|^{2}_2  
\\
      \end{aligned} \label{obj_opt_uncons} 
    \end{gather}
\begin{align}
\mathrm{subject~to:} &  \quad 
\bm{a}(\bm{x})=\bm{J}^{-1}_{\bm{d}}(\bm{x})\underline{\bm{A}}\bm{d}(\bm{x}) \label{diffeo_cnstr}
\end{align}
\end{subequations}
define a solution 
\begin{equation}\label{eq:ABCsol}
    \bm{\psi}=\hat{\bm{d}}^{{[\overline{p}]}}\quad\bm{A}=\bm{A}_{[\overline{p}]}(\underline{\hat{\bm{A}}}),\quad\bm{B}=\hat{\bm{B}},\quad\bm{C}=\hat{\bm{C}}
\end{equation} 
for the optimization problem~\eqref{opt_prob} and thereby define a model of the form \eqref{eq:LTI full}.
\begin{pf}    
For the system class \eqref{eq:sys} satisfying Assumption~\ref{ass:sysCLS}, there exists a smooth equivalence to a linear system without a loss of generality \citep[Theorem 2.3, Corollary 2.1]{Lan2013} as the Koopman-invariance is decoupled from control influence per definition in \eqref{eq:KMDcOp}. By following Proposition \ref{prop1}, we can replace the Koopman-invariance condition \eqref{space_opt} with the constraint \eqref{diffeo_cnstr}. Thus, \eqref{eq:ABCsol} represents a solution for \eqref{opt_prob} - defining a model of the form~\eqref{eq:LTI full}.
\end{pf}
\end{theorem}

The overview of the construction of LTI coordinates can be found in Figure \ref{fig:commDiag}. The unit-box bounded latent space $\mathbb{Y}_{\Box}$ from Fig. \ref{fig:commDiag} is very beneficial for numerical stability.

\begin{remark}
Notably, we are able to learn the autonomous and control part of \eqref{eq:sys} concurrently. Works such as \cite{Korda2020b} necessitate a separation approach - first learning the autonomous and then the control part. This can limit the ability of exploring the state-space from a small amount of initial conditions or become unsafe for certain systems.
\end{remark}

\subsection{Relaxing the Smooth Equivalence through Costs}
To ease the use of standard training algorithms for neural networks, we relax the optimization problem \eqref{eq:step2opt} by considering \eqref{diffeo_cnstr} as a soft constraint through an additional summand in the cost \eqref{obj_opt_uncons}. This results in the unconstrained optimization problem

    \begin{gather}
      \begin{aligned}
\operatornamewithlimits{min}_{\underline{\bm{A}}, \bm{B},\bm{C}, \bm{\bm{d}}(\cdot)}  \sum^{N}_{i=1}&\|\dot{\bm{x}}^{(i)}\tsgn{-}\bm{C}(\bm{A}_{[\overline{p}]}(\underline{\bm{A}})\bm{d}^{{[\overline{p}]}}(\bm{x}^{(i)})+\bm{B}\bm{u}^{(i)})\|^{2}_2\\{+} & \|\bm{x}^{(i)}\tsgn{-}\boldsymbol{C}\bm{d}^{{[\overline{p}]}}(\bm{x}^{(i)})\|^{2}_2+\mathcal{L}_{\text{SE}}(\bm{x}^{(i)}\tsgn{,}\dot{\bm{x}}^{(i)}),\label{obj_opt_unconsFinal} \end{aligned} 
    \end{gather}
where the cost
    \begin{gather}
      \begin{aligned}
\mathcal{L}_{\text{SE}}(\bm{x}^{(i)}, \dot{\bm{x}}^{(i)})=&\|\bm{{\dot{x}}} - \bm{J_{{{\bm{{d_{}}}}}}}^{-1}(\bm{x})\underline{\bm{A}} \bm{{{\bm{{d_{{}}}}}}}(\bm{x})\|_2^2  \label{eq:SE}
\end{aligned} 
    \end{gather}
replaces the constraint \eqref{diffeo_cnstr}. The summands in \eqref{obj_opt_unconsFinal} could also be individually weighted by scalar multipliers to additionally penalize the soft constraints. This allows one to, e.g., selectively bias the learning procedure to prioritize prediction or reconstruction, based on the loss terms defined in \eqref{obj_opt}.

In order to finally solve \eqref{obj_opt_unconsFinal}, one needs to ensure the function approximator used for learning $\bm{d}$ is guaranteed to be a diffeomorphism. For this, we utilize coupling flow invertible neural networks (CF-INN) \citep{45819} which demonstrated their utility for Koopman operator representations \cite{Bevanda2021b, Bevanda2021c}. 

For realizing complex diffeomorphisms, CF-INN successively compose simpler diffeomorphisms called \emph{coupling layers} $\bm{{\hat{d}}_i}$ using the fact that diffeomorphic maps are closed under composition, so that $\bm{y}=\bm{{\hat{d}}}(\bm{x}) = \bm{{\hat{d}}_k} \circ ... \circ \bm{{\hat{d}}_1}(\bm{x})$.
Each coupling layer $\bm{{\hat{d}}_i}$ is defined to couple a disjoint partition of the input $\bm{x}=[\bm{x}^{\top}_a,\bm{x}^{\top}_b]^{\top}$ with two subspaces $\bm{x_a} \in \Set{R}^{d-n}$, $\bm{x_b} \in \Set{R}^n$ where $n \in \Set{N}$ and $d\geq2$, in a manner that ensures bijectivity. This can be realized via affine coupling flows (ACF), which have coupling layers\looseness=-1
\begin{equation}
\bm{{\hat{d}}}_{i}(\bm{x}^{(i)}) = \begin{bmatrix}\bm{x_a}^{(i)}\\\bm{x_b}^{(i)} \odot \operatorname{exp}(\bm{s_i}(\bm{x_a}^{(i)})) + \bm{t_i}(\bm{x_a}^{(i)})\end{bmatrix}
\label{eq:coupling_layer} 
\end{equation}
with scaling functions $\bm{s}_i:\mathbb{R}^n \mapsto \mathbb{R}^{N-n}$ and translation functions $\bm{t}_i:\mathbb{R}^n \mapsto \mathbb{R}^{N-n}$ that can be chosen freely. The parameters of the diffeomorphic learner consist of the the weights and biases in the neural networks of the scaling and translation functions concatenated in parameters $\bm{w}=[\bm{w}^{\top}_{\bm{s}_1},\bm{w}^{\top}_{\bm{t}_1}, \cdots, \bm{w}^{\top}_{\bm{s}_k},\bm{w}^{\top}_{\bm{t}_k}]^{\top}$.

\begin{theorem}\label{thm:GAS}
Consider diffeomorphisms $\bm{d}=\bm{{\hat{d}}_k} \circ ... \circ \bm{{\hat{d}}_1}(\bm{x})$ parameterized through coupling layers  \eqref{eq:coupling_layer}, which are defined using continuously differentiable functions $\bm{s}_i$, $\bm{t}_i$.
Then, by construction, every optimization problem \eqref{obj_opt_unconsFinal} yields a candidate solution for \eqref{eq:step2opt}, resulting in a model of the form \eqref{eq:LTI full}.
\begin{pf}
Following \citep[Appendix - Lemma 11]{Bevanda2021c}, the composition of coupling layers defined using continuously differentiable functions $\bm{s}_i$, $\bm{t}_i$ is guaranteed to be a diffeomorphism. Therefore, any $(\underline{\bm{A}}, \bm{d})$-pair fulfills \eqref{eq:liftedLTI full} by construction, providing the necessary hypothesis class to approximately fulfill \eqref{diffeo_cnstr} by minimizing \eqref{eq:SE}.
\end{pf}
\looseness=-1
\end{theorem}
\begin{remark}
Less formally, minimizing loss \eqref{eq:SE} allows the smooth equivalence defined through $(\underline{\bm{A}}, \bm{d})$ to approximately correspond to the autonomous dynamics of \eqref{eq:sys}.
Thus, when the loss contribution of \eqref{eq:SE} vanishes, the sole error source in the resulting system \eqref{eq:LTI full} is due to a finite truncation of \eqref{eq:KMDcOp}. 

\end{remark}
The result of Theorem \ref{thm:GAS} allows to efficaciously obtain approximate solutions to the optimization problem \eqref{opt_prob}, as it transforms a practically intractable problem \eqref{opt_prob} into an easily implementable deep learning problem of \eqref{obj_opt_unconsFinal}. 

\section{Data-driven LQR for Nonlinear Systems}
\begin{assumption}
The pairs $(\bm{A},\bm{C})$ and $(\bm{A},\bm{B})$ are observable and stabilizable, respectively.
\end{assumption}
The above assumption is common and numerically verifiable for the learned model.
\subsection{Linear Quadratic Optimal Control}
The optimal control problem for \eqref{eq:sys} with quadratic cost is defined as follows
\begin{subequations}
\begin{gather}
\min _{\bm{u}} \int_{t_0=0}^{\infty}\left(\bm{x}(t)^{\top}\bm{Q}\bm{x}(t)+\bm{u}(t)^{\top}\bm{R}\bm{u}(t)\right) dt \label{cost}\\
\text { s.t.} \quad \dot{\bm{x}}=\bm{f}(\bm{x}, \bm{u}), \quad \bm{x}_0=\bm{x}(t_0)
\end{gather}
\end{subequations}

where $\bm{Q} \succeq 0$ and $\bm{R} \succ 0$ are user-defined parameters. Using the solution of \eqref{obj_opt_unconsFinal}, this optimal control problem can be approximated by the following LQR problem
\begin{subequations}\label{eq:KLQR}
\begin{gather}
\min _{\bm{u}} \int_{t_0=0}^{\infty}\left(\bm{z}(t)^{\top}\bm{C}^{\top}\bm{Q}\bm{C}\bm{z}(t)+\bm{u}(t)^{\top}\bm{R}\bm{u}(t)\right) dt\\
\text { s.t.} \quad \dot{\bm{z}}=\bm{A}\bm{z} +\bm{B}\bm{u}, \quad \bm{z}_0=\bm{\psi}(\bm{x}_0)
 \end{gather}
\end{subequations}
by expressing dynamics via the state-space model \eqref{eq:LTI full}. Due to the linear relation of immediate to lifted-state cost, the weight $\bm{C}^{\top}\bm{Q}\bm{C}$ in the resulting optimal control problem \eqref{eq:KLQR} aims at minimizing an equivalent cost functional. This allows for balancing control-energy with aggressiveness in physically relevant coordinates. Its solution delivers a linear feedback gain $\bm{K} \in \mathbb{R}^{m \times D}$ that defines a nonlinear optimal policy in original coordinates $\bm{u}=-\bm{K}\bm{\psi}(\bm{x})$.
{\subsubsection{Optimal Tracking with LQR:}~To admit more challenging tasks such as optimal tracking, we introduce the following reference generating system\!\footnote{$\boldsymbol{x}_r$ is a real vector of arbitrary size, $\bm{r} \in \mathbb{R}^{d}$.}
\begin{subequations}
\label{sysTrack}
\begin{align}
\dot{\bm{x}}_{r} &=\bm{A}_{r} \bm{x}_{r}, \quad \bm{x}_{r}(0)=\bm{x}_{r 0} \label{eq:refDyn} \\
\bm{r} &=\bm{C}_{r} \bm{x}_{r} \label{eq:refOut}
\end{align}
\end{subequations}
such that the cost \eqref{eq:KLQR} is modified for tracking \eqref{eq:refOut} to
\begin{equation}\label{trackCost}
\begin{aligned}
\int_{0}^{\infty}&\left(\left[\boldsymbol{z}^{\top}(t)~\boldsymbol{x}_{r}^{\top}(t)\right]\left[\begin{array}{cc}
\boldsymbol{C}^{\top} \boldsymbol{Q} \boldsymbol{C} & -\boldsymbol{C}^{\top} \boldsymbol{Q} \boldsymbol{C}_{r} \\
-\boldsymbol{C}_{r}^{\top} \boldsymbol{Q} \boldsymbol{C} & \boldsymbol{C}_{r}^{\top} \boldsymbol{Q} \boldsymbol{C}_{r}
\end{array}\right]\left[\begin{array}{c}
\boldsymbol{z}(t) \\
\boldsymbol{x}_{r}(t)
\end{array}\right]\right.\\
&\left.+\boldsymbol{u}^{\top}(t) \boldsymbol{R} \boldsymbol{u}(t)\right) {d} t 
\end{aligned}
\end{equation}
and additionally subject to \eqref{eq:refDyn} - comprising an extended linear system \cite[Section 7]{Lunze2016}.
\begin{equation}\label{sysExt}
    \left[\begin{array}{c}
\dot{\boldsymbol{z}}\\
\dot{\boldsymbol{x}}_{r}
\end{array}\right]=\left[\begin{array}{cc}
\bm{A} & \bm{0} \\
\bm{0} & \bm{A}_{r}
\end{array}\right]\left[\begin{array}{c}
\boldsymbol{z}\\
\boldsymbol{x}_{r}
\end{array}\right]+\left[\begin{array}{c}
\boldsymbol{B}\\
\boldsymbol{0}
\end{array}\right]\bm{u}.
\end{equation}

The tracking LQR law has the form\!\footnote{Using Matlab notation, the solution of the Riccati equation corresponding to \eqref{trackCost} and \eqref{sysExt} is factorized into $\bm{P}=[\bm{P}^{}_{11}, \bm{P}^{}_{12}; \bm{P}^{\top}_{12}, \bm{P}^{}_{22}]$.} $\boldsymbol{u}=-\boldsymbol{R}^{-1} \boldsymbol{B}^{\top} \boldsymbol{P}_{11} \bm{z}-\boldsymbol{R}^{-1} \boldsymbol{B}^{\top} \boldsymbol{P}_{12} \boldsymbol{x}_{r}$. Note that the infinite horizon cost \eqref{trackCost} is only bounded for Hurwitz matrices $\bm{A}_r$. Solutions for the general case are obtained in a finite-horizon fashion.
\section{Evaluation}

For all examples, ACF with 7 coupling layers are used to learn the diffeomorphisms. The neural networks for the scaling and translation functions in each of the affine coupling layers have 3 hidden layers, with 120 neurons, each with a smooth Exponential Linear Unit (ELU) as the activation function. The dimension of the lifting coordinates is $D=65$ ($\bar{p}=10$). As the optimizer, the ADAM \citep{Kingma2015AdamAM} was employed. For Example~\ref{ex:2} both components were learnt jointly from forced data, while for Example~\ref{ex:UUVtoy} the autonomous model and the input matrix were learned separately, with forced and unforced trajectories from the same initial conditions. 
{We discretize the continuous-time state-space matrices $\bm{A}_d$, $\bm{B}_d$ from \eqref{eq:LTI full} for a specified $dt\!=\!0.01$ using the well known closed-form expressions $\bm{A}_{d}=e^{{\bm{A}}dt}$ and $\bm{B}_{d}=\bm{A}^{-1}(\bm{A}_{d}-\bm{I})\bm{B}$. The weights are chosen to be $\bm{Q} = 100\bm{I}$, $\bm{R} = 0.01$. In optimal tracking, the finite-horizon is chosen to be $T=50s$. In our experiments, we use the discrete-time counterpart of the optimal tracking LQR for convenience. The terminal weight is chosen as $\bm{P}(N)= \bm{Q}+\bm{Q}^{\top}$, with $N=T/dt$.}
\begin{example}\label{ex:UUVtoy}
Consider the following stable dynamical system containing characteristic hydrodynamic damping of underwater vehicles \citep{bookFOSS}
\begin{equation}\label{SMdt}
\bm{{\dot{x}}}=\left[\begin{array}{l}
\dot{x}_{1} \\
\dot{x}_{2}
\end{array}\right]=\left[\begin{array}{c}
x_{2} \\
-x_{1}-x_2-x_{2}\left|x_{2}\right|+u
\end{array}\right] .
	\end{equation}
	 We gather trajectory data from 50 uniformly spaced initial conditions, starting from the edge of $[-2.5,2.5]^2$ for $5s$ with sampling time $dt=0.025s$, while control inputs are generated from an amplitude modulated pseudo-random bit sequences (APRBS) signal with amplitude from $-1$ to $1$ and hold time from $0.025s$ to $0.1s$, resulting in 200 datapoints each. The neural network is trained for 10000 epochs with a full-batch gradient descent. {For optimal tracking, a circular trajectory is prescribed by choosing marginally stable $\bm{A}_r$ and $\bm{C}_r=\bm{I}$.
	 Learning the nonlinear effects decreases the accumulated cost compared to an exact first order Taylor linearization around the equilibrium. Figure \ref{fig:my_label1} shows how our proposed KFc-LQR leads to drastically improved tracking - reducing the accumulated $L_2$ tracking error by $97.06\%$.}
\end{example}

%
\looseness=-1
\begin{figure}[ht!]
    \centering
    \setlength\fheight{0.8\textheight} 
    \setlength\fwidth{0.66\textwidth}
%
%
\begin{tikzpicture}

\begin{axis}[%
width=0.6\fwidth,
height=0.1\fheight,
at={(0\fwidth,0.15\fheight)},
scale only axis,
xmin=0,
xmax=40,
xlabel style={font=\color{white!15!black}},
ymin=-5,
ymax=5,
ylabel style={font=\color{white!15!black}},
ylabel={${ x}_\text{1}$},
axis background/.style={fill=white},
xmajorgrids,
ymajorgrids,
]
\addplot [forget plot,color=blue, line width=1.5pt]
  table {y0.dat};
\addplot [forget plot,color=red, line width=1.5pt]
  table {y0tl.dat};
\addplot [forget plot,color=black, dotted, line width=1.5pt]
  table {y0ref1.dat};

\end{axis}

\begin{axis}[%
width=0.6\fwidth,
height=0.1\fheight,
at={(0\fwidth,0\fheight)},
scale only axis,
xmin=0,
xmax=40,
xlabel style={font=\color{white!15!black}},
ymin=-4,
ymax=6,
ylabel style={font=\color{white!15!black}},
ylabel={${ x}_\text{2}$},
xlabel={$t$},
axis background/.style={fill=white},
xmajorgrids,
ymajorgrids,
    legend style={draw=white!15!black,legend cell align=left,nodes={scale=0.8}},
    legend columns=3,
]
\addplot [forget plot,color=blue, line width=1.5pt]
  table {y1.dat};
\addplot [forget plot,color=red, line width=1.5pt]
  table {y1tl.dat};
\addplot [forget plot,color=black, dotted, line width=1.5pt]
  table {y1ref1.dat};
      \addlegendimage{blue, thick}
    \addlegendentry{KFc-LQR};
    \addlegendimage{red, thick}
    \addlegendentry{LQR};
        \addlegendimage{black, dotted, thick}
    \addlegendentry{reference};
\end{axis}


\pgfplotsset{
	legend cell align = left,
	legend style = {font=\tiny, fill opacity=1.0},
	title style={yshift=-7pt, font = \small} }
\makeatletter
\tikzset{
    dot diameter/.store in=\dot@diameter,
    dot diameter=3pt,
    dot spacing/.store in=\dot@spacing,
    dot spacing=3pt,
    dots/.style={
        line width=\dot@diameter,
        line cap=round,
        dash pattern=on 0pt off \dot@spacing
    }
}

\end{tikzpicture}%
    \caption{A demonstration of superior tracking performance of KFc-LQR for Example \ref{ex:UUVtoy}}
    \label{fig:my_label1}
\end{figure}
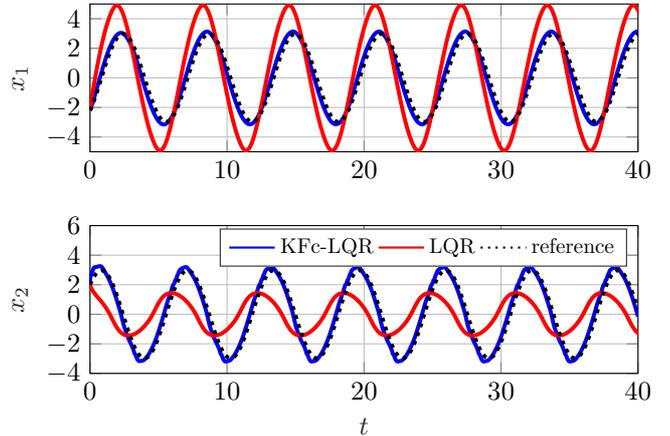
\looseness=-1
\begin{example}\label{ex:2}
Here we consider control of the following system
\begin{equation}
\bm{{\dot{x}}}=\left[\begin{array}{c}
x_{2} \\
-5x_1-0.3x_2-v(x_2)-t(x_1,x_2) + u
\end{array}\right]
	\end{equation}
with velocity-related nonlinear damping $v(x_2)=5(x_{2}^3+x_{2}\left|x_{2}\right|)$ as well as complex damping effects $t(x_1,x_2)=10x_2\sin(5x_1)\cos (2x_1)$ dependent on the position.		

For this example, 1024 uniformly spaced initial conditions are sampled from the set $[-1,1]^2$ for $0.5s$ ($dt=0.005s$) resulting in 100 datapoints each while controlled using an APRBS signal with amplitude from $-5$ to $5$ and hold time $0.01s$ to $0.1s$. The neural network is trained for 20 000 epochs with a batch size of 2000. 
{
For a comparison to established data-driven methods, we take the continuous-time framework of gEDMD \citep{Klus2020} extended for controls as in to \cite{Korda2018a} - gEDMDc. We have tested monomials and radial basis functions as a basis for gEDMDc and sided with monomials as they performed better. For optimal tracking, a sinusoidal trajectory in $x_1$ is prescribed by choosing a marginally stable $\bm{A}_r$ and $\bm{C}_r=[1~0]$. KFc-LQR reduces the accumulated $L_2$~tracking error by $80.34\%$ for the same model order ($D\!=\!65$) as for gEDMDc. The much improved tracking performance is supported by a more adequate control policy and shown in Figure \ref{fig:my_label2}. This shows the importance of learning LTI features and dynamics jointly. Due to the features being system-specific with non-existing apriori intuition, predefining them as in gEDMDc is limiting.}
\end{example}
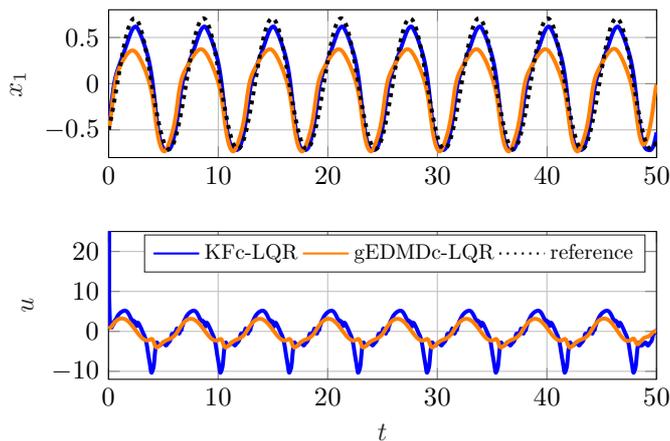
\begin{figure}[ht!]
    \centering
    \setlength\fheight{0.8\textheight} 
    \setlength\fwidth{0.66\textwidth}
%
%
\begin{tikzpicture}

\begin{axis}[%
width=0.6\fwidth,
height=0.1\fheight,
at={(0\fwidth,0.15\fheight)},
scale only axis,
xmin=0,
xmax=50,
xlabel style={font=\color{white!15!black}},
ymin=-0.8,
ymax=0.8,
ylabel style={font=\color{white!15!black}},
ylabel={${ x}_\text{1}$},
axis background/.style={fill=white},
xmajorgrids,
ymajorgrids,
]
\addplot [forget plot,color=blue, line width=1.5pt]
  table {y0e2.dat};
\addplot [forget plot,color=orange, line width=1.5pt]
  table {y0gE.dat};
\addplot [forget plot,color=black, dotted, line width=1.5pt]
  table {y0ref2.dat};
\end{axis}
\begin{axis}[%
width=0.6\fwidth,
height=0.1\fheight,
at={(0\fwidth,-0.0\fheight)},
scale only axis,
xmin=0,
xmax=50,
xlabel style={font=\color{white!15!black}},
ymin=-12,
ymax=25,
ylabel style={font=\color{white!15!black}},
ylabel={${u}$},
xlabel={$t$},
axis background/.style={fill=white},
xmajorgrids,
ymajorgrids,
    legend style={draw=white!15!black,legend cell align=left,nodes={scale=0.8}},
    legend columns=3,
]
\addplot [forget plot,color=blue, line width=1.5pt]
  table {u2kf.dat};
\addplot [forget plot,color=orange, line width=1.5pt]
  table {u2ge.dat};
      \addlegendimage{blue, thick}
    \addlegendentry{KFc-LQR};
    \addlegendimage{orange, thick}
    \addlegendentry{gEDMDc-LQR};
        \addlegendimage{black, dotted, thick}
    \addlegendentry{reference};
\end{axis}


\pgfplotsset{
	legend cell align = left,
	legend style = {font=\tiny, fill opacity=1.0},
	title style={yshift=-7pt, font = \small} }
\makeatletter
\tikzset{
    dot diameter/.store in=\dot@diameter,
    dot diameter=3pt,
    dot spacing/.store in=\dot@spacing,
    dot spacing=3pt,
    dots/.style={
        line width=\dot@diameter,
        line cap=round,
        dash pattern=on 0pt off \dot@spacing
    }
}

\end{tikzpicture}%
    \caption{Superior tracking performance compared to gEDMDc-LQR though the use of KFc-LQR in Example \ref{ex:2}}
    \label{fig:my_label2}
\end{figure}
\looseness=-1
\section{Conclusion}
In this paper, we present a novel framework for learning Koopman generator models for controlled systems by constructing controlled LTI prediction models for a class of nonlinear dynamics. 
{We demonstrate that our KFc-LQR framework delivers superior model-based control performance compared to established and related methods for LQR control based on LTI models.} Future work considerations include rigorous analysis of controllability and observability form data and extending the model capacity to that of systems with state-dependent actuation.}
\begin{ack}
We thank Jan Br\"{u}digam for useful insights in the preparation this manuscript.
\end{ack}
\bibliography{ifacconf}             
\end{document}